\documentclass[fleqn,usenatbib]{mnras}
\usepackage{multicol}
\usepackage{graphicx}
\usepackage{subcaption}
\usepackage{newtxtext,newtxmath}
\usepackage{url}
\usepackage[T1]{fontenc}

\DeclareRobustCommand{\VAN}[3]{#2}
\let\VANthebibliography\thebibliography
\def\thebibliography{\DeclareRobustCommand{\VAN}[3]{##3}\VANthebibliography}

\usepackage{graphicx}	
\usepackage{amsmath}	

\title[Modelling JKCS041]{Decoding the Early Universe: Exploring a Merger Scenario for the High-Redshift Cluster JKCS041 using Numerical Models}

\author[Sharon Felix et al.]{
Sharon Felix,$^{1,2}$\thanks{E-mail:sxf18000@utdallas.edu}
Antareep Gogoi,$^{1}$
Kaitlyn Shavelle,$^{3,4}$
Brandon Sike,$^{1,5}$
Lindsay King,$^{1}$
Stefano Andreon,$^{6}$ \newauthor
Urmila Chadayammuri,$^{7}$
John ZuHone,$^{7}$
Charles Romero$^{7}$
\\
\\
$^{1}$Department of Physics, University of Texas at Dallas, 800 W Campbell Rd, Richardson, TX 75080, USA\\
$^{2}$Department of Physics, Austin College, 900 N Grand Ave, Sherman, TX 75090, USA\\
$^{3}$Department of Astronomy, Columbia University, New York, NY 10027, USA\\
$^{4}$ Maria Mitchell Observatory, 3 Vestal St, Nantucket, MA 02554, USA\\
$^{5}$Department of Astronomy, University of Michigan, 500 S State St, Ann Arbor, MI 48109, USA\\
$^{6}$INAF–Osservatorio Astronomico di Brera, via Brera 28, 20121, Milano, Italy\\
$^{7}$Center for Astrophysics $\vert$ Harvard \& Smithsonian, 60 Garden Street, Cambridge, MA 02138, USA
}

\date{Accepted XXX. Received YYY; in original form ZZZ}

\pubyear{2023}

\begin{document}
\label{firstpage}
\pagerange{\pageref{firstpage}--\pageref{lastpage}}
\maketitle

\begin{abstract}
JKCS041 ($z=1.8$) is one of the most distant galaxy cluster systems known, seen when the Universe was less than 4 billion years old. Recent Sunyaev-Zeldovich (SZ) observations show a temperature decrement that is less than expected based on mass estimates of the system from X-ray, weak gravitational lensing and galaxy richness measurements. In this paper we seek to explain the observables - in particular the low SZ decrement and single SZ peak, the projected offset between the X-ray and SZ peaks of $\approx$220\,kpc, the gas mass measurements and the lensing mass estimate. We use the GAMER-2 hydrodynamic code to carry out idealized numerical simulations of cluster mergers and compare resulting synthetic maps with the observational data. The observations are not well reproduced by an isolated cluster, while instead they are when considering cluster mergers viewed a few tenths of a Gyr after first core passage. A range of merger scenarios is consistent with the observations, but parts of parameter space can be ruled out, and generically some kind of merger process is necessary to reproduce the offset between the SZ and X-ray peaks. In particular, a total mass of $\approx$2$\times 10^{14} M_\odot$, mass ratio of $\approx$2:3, gas fraction of $0.05-0.1$ and Navarro, Frenk and White (NFW) mass density profile concentration $c$$\approx$5 for both components are scenarios that are consistent with the observational data. 
\end{abstract}


\section{Introduction}
 Galaxy clusters are the most massive bound objects in the Universe, assembling hierarchically over time from galaxies, groups and lower mass clusters. Most of the matter in the Universe is dark \citep{1957moas.book.....Z}, making up about 80\% of the mass in most clusters \citep{1999fsu..conf.....D}. The luminous matter is primarily ionised gas, which emits X-rays and can also be detected via a Sunyaev-Zel'dovich (SZ) signal against the Cosmic Microwave Background (CMB) \citep{1969Ap&SS...4..301Z, 1970Ap&SS...7....3S}. Stars that light up galaxies account for only a small fraction of the luminous matter, e.g.,  \citet{2010MNRAS.407..263A}. Only a few clusters at $z\ge 1.75$ have detected intracluster medium (ICM); these are XLSSC 122 \citep{2018A&A...620A...2M}, IDCS J1426 \citep{2012ApJ...753..164S}
and  JKCS041 \citep{2009A&A...507..147A}.

Galaxy cluster mergers are among the largest dynamic events in the Universe, occurring when clusters collide under gravity. They exhibit detectable features in the X-ray emitting ICM such as shocks and turbulence \citep{2007PhR...443....1M, 2019SSRv..215...24S, 1995A&A...302L...9B, 1998ApJ...493...62R, 1998ApJ...503...77M}. Seen in the aftermath of a collision, the ``Bullet cluster" \citep{2004ApJ...604..596C} provides the first example of the merger process resulting in plasma clouds lagging behind the dark matter and galaxy contents of the individual clusters. Numerical simulations are important tools to study the cluster merger processes that lead to the observed features \citep{2009ApJ...699.1004Z, 2011ApJ...728...54Z}. In our hierarchical Universe, mergers are common at $z\sim2$ in large-volume cosmological simulations of structure formation (see Table 2 of the review \cite{2020Vog}), a time at which we expect
many clusters to be forming, rather than fully evolved. 

JKCS041 has been the target of multi-wavelength observations including X-ray (\textit{Chandra X-ray Observatory} \citep{2000SPIE.4012....2W}) and SZ (MUSTANG-2 \citep{2020ApJ...902..144D,2014JLTP..176..808D} on Green Bank Telescope) measurements. The cluster system was detected in 2006 using \textit{J} and \textit{K} data from the UKIRT (United Kingdom Infrared Telescope) Infrared Deep Sky Survey (UKIDSS, \citet{2007MNRAS.379.1599L}) Early Data Release \citep{2006MNRAS.372.1227D}, by applying the red sequence method developed by \citet{2003A&A...409...37A}. \citet{2009A&A...507..147A} obtained a photometric redshift estimate of 1.90. \citet{2014ApJ...788...51N} and \citet{2014A&A...565A.120A} determined a redshift $z=1.803$ using \textit{Hubble Space Telescope} grism spectroscopy, based on 19 galaxies that are confirmed cluster members. The mass of the cluster was later estimated to be $ \log(M/M_\odot) \ge$ 14.2 from gas mass, X-ray luminosity and X-ray temperature studies; this massive system, seen when the Universe was about 3.6 billion years old, is most likely a progenitor of a cluster like the Coma cluster that we see today \citep{2014A&A...565A.120A}. 

In this work, we obtain computational models for the dark matter and plasma in JKCS041 that reproduce the main observational features. In particular, we make synthetic measurements using SZ and X-ray maps created from merging clusters in numerical simulations, accounting for observational factors such as projection effects and instrumental responses which are present in real data. With one of the best fit mass models for the system, constrained by SZ and X-ray observations, we then carry out a synthetic weak lensing analysis using a distant galaxy source population, consistent with the recent work of \citet{kim}. 

In Section \ref{sec:2}, we outline the primary observables that are used to constrain the models. In Section \ref{sec:3}, idealized numerical simulations of the system and the parametric models used for the study of merging clusters are described. We model JKCS041 as a cluster merger comparing the simulations with the observed quantities in Section \ref{sec:5}. Note that we harness the flexibility of idealized simulations, rather than cosmological simulations, to explore the parameter space for this rare cluster system.  In Section \ref{sec:6} we outline a gravitational lensing analysis, using our best fit model from Section \ref{sec:5} as a lens, and comparing the mass estimates with the lensing analysis of \citet{kim}. In Section \ref{sec:7}, we discuss the results of our study and some of the caveats of the simulations, and present the conclusions.

Throughout this paper we assume a flat Universe with cosmological parameters $\Omega_{\rm m}=0.3$, $\Omega_{\Lambda}=0.7$ and present day Hubble constant $H_{0}=70 \, \rm km \, \rm s^{-1} \, \rm Mpc^{-1}$. 

\section{Observational constraints on JKCS041}
\label{sec:2}
In this section we outline the observational constraints on JKCS041 that will be used to construct numerical models, noting the features that we seek to model.

JKCS041 was observed by \textit{Chandra} using ACIS-S for 75\,ks \citep{2009A&A...507..147A}
and by GBT using MUSTANG-2 for 28 hours \citep{2023MNRAS.522.4301A}. The SZ map shows a single
temperature decrement peak, centred close to the BCG and offset by $\approx$220\,kpc North of the
X-ray peak \citep{2023MNRAS.522.4301A}. To account for the ambiguity in defining the SZ and X-ray centres, in this paper we allow for $10 \%$ uncertainty on this offset. The observed value of SZ signal averaged inside a $10''$ radius around the peak is $62^{+12}_{-12} \mu K$ \citep{2023MNRAS.522.4301A}. For comparison, in our simulations we average the SZ temperature decrement inside the same radius; hereafter we refer to this averaged quantity as the peak SZ temperature decrement ($\Delta T$).
The X-ray profile core radius, $r_{c}$, was found to be $36^{+8.3}_{-7.6}$ arcseconds \citep{2009A&A...507..147A}, which is $\approx$310\,kpc at $z=1.8$ and is used for all the simulations in this paper. 
The gas mass inside $30^{''}$, $40^{''}$ and $60^{''}$ is found to be $3.9 (\pm 0.8) \times 10^{12} M_\odot$, $7.7 (\pm 1.6) \times 10^{12} M_\odot$ and $1.8 (\pm 0.4) \times 10^{13} M_\odot$ respectively \citet{2009A&A...507..147A, 2014A&A...565A.120A}. In a weak lensing analysis of deep data, \citet{kim}, found $M_{200}=4.7 (\pm 1.5) \times 10^{14} M_\odot$ for a single component NFW profile fit \citep{1996ApJ...462..563N, 1997ApJ...490..493N}.

In summary, the observational features serving as constraints for the numerical simulations are: (i) the single intense SZ peak; (ii) the magnitude of SZ signal; (iii) observed SZ-X-ray offset; (iv) gas mass profile. The resulting model is further compared with the lensing map of \citet{kim}. 

\section{Numerical Simulations}
\label{sec:3}
In this section we introduce the numerical simulations of cluster systems that are compared with the observational constraints presented in Section \ref{sec:2}.
We use GAMER-2 (GPU-Accelerated Adaptive MEsh Refinement) \citep{2010ApJS..186..457S, 2018MNRAS.481.4815S,2018ApJS..236...50Z} to study the observational signatures of isolated and merging clusters, including synthetic measurements of SZ and X-ray maps. The initial conditions for the clusters are generated using the \hyperlink{https://github.com/jzuhone/cluster_generator}{cluster generator package} \footnote{https://github.com/jzuhone/cluster\_generator}. To describe merging clusters, we use spherical dark matter haloes hosting X-ray emitting plasma clouds parameterized as described in Section \ref{sec:3.1}. For mergers, additional parameters such as initial relative velocities and impact parameter for the clusters are necessary, as described in Section \ref{sec:5}. 

GAMER-2 is an adaptive mesh refinement (AMR) code, where the spatial and temporal grid on which we solve the dynamic equations of motion adaptively adjust so that local regions requiring higher resolution are identified and given more computational resources. GPU acceleration reduces the computation time without sacrificing accuracy, as studied by \cite{2018MNRAS.481.4815S}. The code uses gas entropy per volume to calculate the pressure and temperature given by 
\begin{equation}
    s=\frac{P}{\rho^{\gamma-1}}
\end{equation}
where P is the pressure, $\rho$ is the gas mass density and $\gamma$ is the gas adiabatic index \citep{2018MNRAS.481.4815S}. 

Given boundary conditions (in this case isolated) the gravitational potential is evaluated by solving the discretized Poisson equation. The simulations that we carried out do not account for radiative transfer, cooling, cluster galaxies or large-scale structure. 

\subsection{Dark matter and gas profiles}
\label{sec:3.1}

The cluster-scale dark matter haloes in the simulation are described by the super NFW (sNFW) profile \citep{2018MNRAS.476.2086L, 2018MNRAS.478.1281L}, a model very similar to NFW \citep{1996ApJ...462..563N, 1997ApJ...490..493N} but with finite mass ensuring a smooth cutoff at larger radii. The equations for the sNFW profile given below are taken from \citet{2018MNRAS.476.2086L}. NFW haloes are usually parameterized by a mass $M_{200}$ or radius $r_{200}$ and mass concentration parameter $c$. $M_{200}$ is the mass contained inside radius $r_{200}$, at which the mean enclosed density is 200$\rho_{\rm crit}$. The critical density, $\rho_{\rm crit} = 3H(z)^2/8\pi G$, where $H(z)$ is the Hubble parameter at the redshift $z$. The NFW scale radius is defined as $r_s = r_{200}/c$. The sNFW density profile is given by:

\begin{equation}
   \rho_{snfw}(r) =\frac{3M_s}{16 \pi a^{3}} \frac{1}{(\frac{r}{a})(1+\frac{r}{a})^{\frac{5}{2}}}\,,
   \label{eq8}
\end{equation}

where $a$ is the sNFW scale radius which is related to the half-mass radius $R_{e}$ by $a=R_{e}/5.478$. The $M_s$ parameter is determined by $a$ and an sNFW concentration parameter $c_{sNFW}$. Fitting sNFW haloes with NFW density profiles, \citet{2018MNRAS.476.2086L} found that the concentrations $c_{sNFW}$ and $c$ are related by: 
\begin{equation}
    c_{sNFW}=1.36+0.76c\,.
    \label{csnfw}
\end{equation} 
The scale radius of an NFW profile $r_{s}$ is related to the sNFW $a$ parameter by \citet{2018MNRAS.476.2086L}:
\begin{equation}
    r_{s}=\frac{2a}{3}
    \label{rsc}
\end{equation}
Essentially, the parameters we study are NFW parameters $M_{200}$ and $c$ that map to the sNFW profile (equation \ref{eq8}) parameters $M_s$ and $a$ as described in \citet{2018MNRAS.476.2086L}.

We set up the initial clusters with gas density following the modified $\beta$-profile as described in \citep{2006ApJ...640..691V}, along with requiring that the gas and dark matter are in  hydrostatic equilibrium. This assumption is valid only for relaxed clusters and will be discussed in Section \ref{sec:7}. This is achieved by giving the dark matter particles velocities after the mass profiles are set, as per \citet{2004ApJ...601...37K} in which the energy distribution is calculated using the Eddington formula \citep{1916MNRAS..76..572E}.

Most clusters in the local Universe are characterized by a peak in the X-ray surface brightness and a lower temperature in the centre denoting a `cool core'. The distribution of the gas in cool core clusters is described by a modified-$\beta$-profile \citep{2006ApJ...640..691V}, 
\begin{equation}
    n_{p}n_{e}=n_{0}^{2} \frac{(r/r_{c})^{-\alpha}}{(1+r^{2}/r_{c}^{2})^{3\beta-\alpha/2}} \frac{1}{(1+r^{\gamma}/r_{s}^{\gamma})^{\epsilon/\gamma}}\,,
    \label{gas_prof}
\end{equation}
 where $n_e$ and $n_{p}$ are the electron and proton number densities. $\alpha $ is the inner density slope which determines the strength of the cool core and $r_c$ is the core radius. Cool cores are well-represented by parameters similar to $r_{c}=0.05a$, $r_{s}=0.6a$ and $\alpha=2$ \citep{2022MNRAS.509.1201C}. We use $\beta = 2/3$ which was found to fit X-ray surface brightness of clusters in \citet{1984ApJ...276...38J} and \citet{2006ApJ...640..691V}. The parameter $\epsilon$ describes a steepening of slope near the radius $r_{s}$ and the parameter $\gamma$ controls the width of the transition. The default values that will be adopted for the remaining plasma cloud parameters are $\gamma=3$ and $\epsilon=2$, consistent with studies of multiple clusters in \citet{2006ApJ...640..691V}. 
 
 The initial conditions for the merger simulations consist of specification of sNFW parameters (using NFW parameters, $M_{200}$ and $c$ through equations \eqref{rsc} and \eqref{csnfw} and gas parameters for the two haloes, initial distance between the haloes, impact parameter and the initial halo velocities (or relative velocity of the haloes). 
 
\begin{figure}
    \centering
\includegraphics[width=.4 \textwidth]{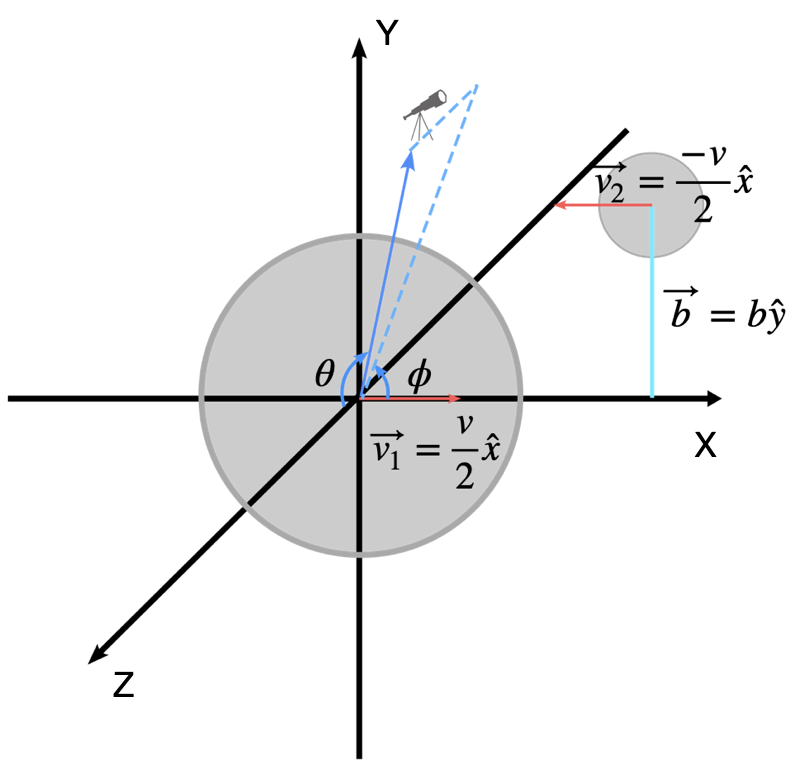}
    \caption{The merger geometry. The velocities are in the X-direction and the impact parameter is set in the Y-direction. The merger process happens in the X-Y plane. The polar angle $\theta$ and the azimuthal angle $\phi$ describe the direction of an observer viewing the merger at a viewing direction indicated by the telescope. Unless otherwise stated, the projections will be assumed to be on the Z-axis, equivalent to the observer being on the Z-axis.}
    \label{geometry}
\end{figure}

 Fig.\,\ref{geometry} shows the geometry of the merging process, and location of the observer. The simulation box is a cube of 14 Mpc side length where the objects are placed 3 Mpc apart initially. The individual clusters are then given an initial velocity towards each other along the X-axis and hence the merger happens primarily along the X-axis. The impact parameter specifies the displacement in the Y direction of the second cluster from the origin at the start of the simulation. Unless otherwise specified, we use the Z-axis projection to study the evolution of the merger with time (equivalent to the observer being on the Z-axis). We run the simulation for 2.5 Gyr, which allows the system to evolve for sufficient time after first pericentre passage in all of our runs. In the highest mass configuration we ran, this time was enough for a second pericentre passage. Considering the age of the Universe at $z=1.803$ ($\approx$3.6\,Gyr), a longer time evolution is not favored. 
 
 For ease of analysis, for the merger scenario we analyze each of the configurations with data outputs between 1.2 Gyr and 2.5 Gyr at a frequency of every 0.1 Gyr. This time window is chosen as it takes approximately 1.2 Gyr from the start of simulation for the clusters to get sufficiently close so that the behavior starts to deviate from that of isolated clusters. This time range also gives us enough pre-pericentre passage data.

 For the main set of merger simulations that we run, and considering the masses we use, pericentre passage occurs between 1.5 Gyr and 1.8 Gyr after the start of the simulation. Therefore, we choose the initial redshift as $z=3.4$ which corresponds approximately to $1.8$ Gyr before the observed system redshift $z=1.8$. 

We use the output of the GAMER-2 \citep{2018MNRAS.481.4815S, 2018ApJS..236...50Z} simulations in combination with the yt code \citep{2011ApJS..192....9T,yt.astro.analysis} to obtain synthetic SZ, X-ray and gravitational lensing maps.
 
\section{Modelling JKCS041 using SZ and X-ray mock observations}
\label{sec:5}

Our main goals for modelling JKCS041 using numerical simulations are to explain the SZ signal observed (single peak and low decrement), the X-ray gas mass measurements, and the offset between the SZ and X-ray peaks. As previously noted, the observer is assumed to be on the Z-axis (and projections along the Z-axis) unless otherwise stated. Prior to our main analysis, we considered isolated spherical single clusters but determined that these cannot reproduce the offset observed between SZ and X-ray peaks. In addition, a merger scenario is indicated by the strength of the SZ signal in comparison with the estimated cluster mass as discussed in Section \ref{sec:5.1}. Hence we consider merger scenarios for our analysis. 

Table \ref{Table:1} describes the run configurations studied in this work, motivated by various preliminary runs that were carried out to assess the impact of merger parameters on observables. 
We divide the simulations into a grid, within which simulations vary a single quantity, namely: total mass, concentration parameter, mass ratio, impact parameter, and gas fraction. In these runs the initial relative velocity is set at 1100 km/s for all of the systems, consistent with the velocities of clusters in cosmological simulations, and such that the lowest mass systems remain dynamically bound. Later in section \ref{sec:4.2}, we comment on the impact of initial relative velocity and of the viewing angles ($\theta$ and $\phi$). At each time step, we 
generate SZ temperature, Compton y, X-ray emissivity, projected density maps etc. of the simulation data. 
Simulated Compton y maps are convolved with the beam and the transfer function of the real Green Bank Telescope MUSTANG-2 data (from \citealp{2023MNRAS.522.4301A}) to compare them with the observed map. We use the APEC model for calculation of X-ray emissivity through post processing of the simulation output \citep{2001ApJ...556L..91S} and an energy range consistent with the \emph{Chandra} observations.

We define ${f_\text{gas}}$ as the ratio of gas mass to total cluster mass inside $r_{200}$.
\setlength{\arrayrulewidth}{0.2mm}
\setlength{\tabcolsep}{12pt}
\renewcommand{\arraystretch}{1.2}
\begin{table*}
\centering
\begin{tabular}{ |p{.5cm}|p{.3cm}|p{.5cm}| p{.5cm} |p{.3cm}| p{.5cm} |p{.3cm}|p{.8cm} }
\hline
Run & $c_{1}$, $c_{2}$ & $M_{1}$ ($10^{14}\ M_\odot$) & $M_{2}$ ($10^{14}\ M_\odot$) & q & ${f}_{\rm gas}$& I (kpc)& $\Delta T_{\rm avg} \ (\mu K)$ \\
\hline
\multicolumn{8}{|c|}{Var$_{M}$  : Changing total mass}\\
\hline
$\romannumeral 1$& 3 & 3 & 3 &1:1& 0.05 & 0 & -260\\
$\romannumeral 2$& 3 & 2 & 2 & 1:1& 0.05 & 0&-149\\
$\romannumeral 3$& 3 & 1.5 & 1.5 & 1:1& 0.05& 0&-106\\
$\romannumeral 4$& 3 & 1.2 & 1.2 & 1:1& 0.05& 0&-78\\
$\romannumeral 5$& 3 & 1.0 & 1.0 & 1:1& 0.05& 0&-69\\
\hline
\multicolumn{8}{|c|}{Var$_{c}$: Changing concentration parameter}\\
\hline
$\romannumeral 6$& 3 & 1.0 & 1.0 & 1:1& 0.05& 0&-69\\
$\romannumeral 7$& 4 & 1.0 & 1.0 & 1:1& 0.05& 0&-66\\
$\romannumeral 8$& 5 & 1.0 & 1.0 & 1:1& 0.05& 0&-72\\
\hline
\multicolumn{8}{|c|}{Var$_{q}$  : Changing mass ratio}\\
\hline
$\romannumeral 9$& 5 & 1.0 & 1.0 & 1:1& 0.05& 0&-72\\
$\romannumeral 10$& 5 & 0.5 & 1.5 & 1:3& 0.05& 0&-61\\
$\romannumeral 11$& 5 & 0.8 & 1.2 & 2:3& 0.05& 0&-61\\
\hline
\multicolumn{8}{|c|}{Var$_{I}$ : Changing impact parameter}\\
\hline
$\romannumeral 12$& 5 & 0.8 & 1.2 & 2:3& 0.05& 0&-61\\
$\romannumeral 13$& 5 & 0.8 & 1.2 & 2:3& 0.05& 100&-62\\
$\romannumeral 14$& 5 & 0.8 & 1.2 & 2:3& 0.05& 500&-51\\
\hline
\multicolumn{8}{|c|}{Var$_{{f}_{\rm gas}}$: Changing gas fraction}\\
\hline
$\romannumeral 15$& 5 & 0.8 & 1.2 & 2:3& 0.05& 0&-61\\
$\romannumeral 16$& 5 & 0.8 & 1.2 & 2:3& 0.10& 0&-72\\
$\romannumeral 17$& 5 & 0.8 & 1.2 & 2:3& 0.17& 0&-130\\
\hline
\multicolumn{8}{|c|}{Best fit}\\
\hline
$\romannumeral 18$& 5 & 0.8 & 1.2 & 2:3& 0.05& 0&-61\\
\hline

\end{tabular}
\caption{Set of simulations carried out to model JKCS041 as a cluster merger. The mass concentration parameters are denoted by $c_{1}$ and $c_{2}$, masses by $M_{1}$, $M_{2}$, mass ratio $q$, gas fraction ${f}_{\rm gas}$, impact parameter $I$ and $r_{c}$ gas core radius. The $\Delta T_{\rm avg}$ denotes the average $\Delta T$ within a $10''$ around the peak value. Throughout these runs we use an initial relative velocity of 1100 km/s; we return to this later in Sec. \ref{sec:5.2} }.
\label{Table:1}
\end{table*}

\subsection{General trends of the SZ signal during a merger}
\label{sec:5.1}

\begin{figure}
    \centering
    \includegraphics[width=.45 \textwidth]{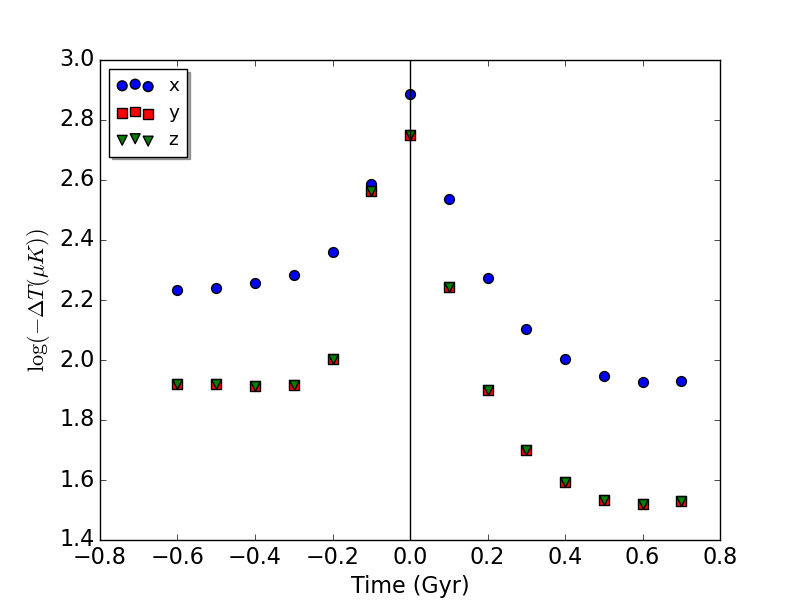}
    \caption{Logarithm of the peak absolute value of SZ $\Delta T$ as a function of time, measured from the beginning of the simulation for a typical run (
    run $\romannumeral 5$
    in Table \ref{Table:1}).
    Different projections are shown, as detailed in the legend.
    The SZ signal has a peak at the pericentre passage, indicated by a solid line.}
    
    \label{fig:11}
\end{figure}

First, we explore the evolution of the SZ peak (as defined in Section \ref{sec:2}) during a typical merger configuration. Fig.\,\ref{fig:11} shows the variation of the SZ peak with time for a typical configuration (run $\romannumeral 5$ in Table\,\ref{Table:1}). When we refer to SZ peak values, we implicitly mean $|\Delta T|$ peak values, which are in practice directly obtained from SZ observations. At the start of the simulations, the SZ peak remains almost constant until the clusters are sufficiently close together, when it increases to reach a maximum around the pericentre and then decreases after that. After pericentre passage the SZ peak attains a lower value than at the start because of the lower gravitational binding of the gas making the distribution more diffuse compared to the distribution before pericentre passage. Therefore, a merger can produce lower SZ peak than the SZ peak of the cluster components in isolation. These trends for mergers are consistent with  \citet{2008ApJ...680...17W} and \citet{2012MNRAS.419.1766K}. 

\subsection{Constraining the total mass with SZ observations}
\label{sec:4.2}

In order to narrow down parameter space of mergers that are consistent with the observables, in this section we start by considering the impact of total mass (first, fixing mass ratio $q=1$) on SZ maps. We then constrain the concentration $c$, mass ratio $q$, Impact parameter $I$ and gas fraction ${f}_{\rm gas}$. This process allows us to examine the sensitivity of the SZ signal to these parameters and rule out regions of parameter space.

\begin{figure}
    \centering
\includegraphics[width=.48\textwidth]{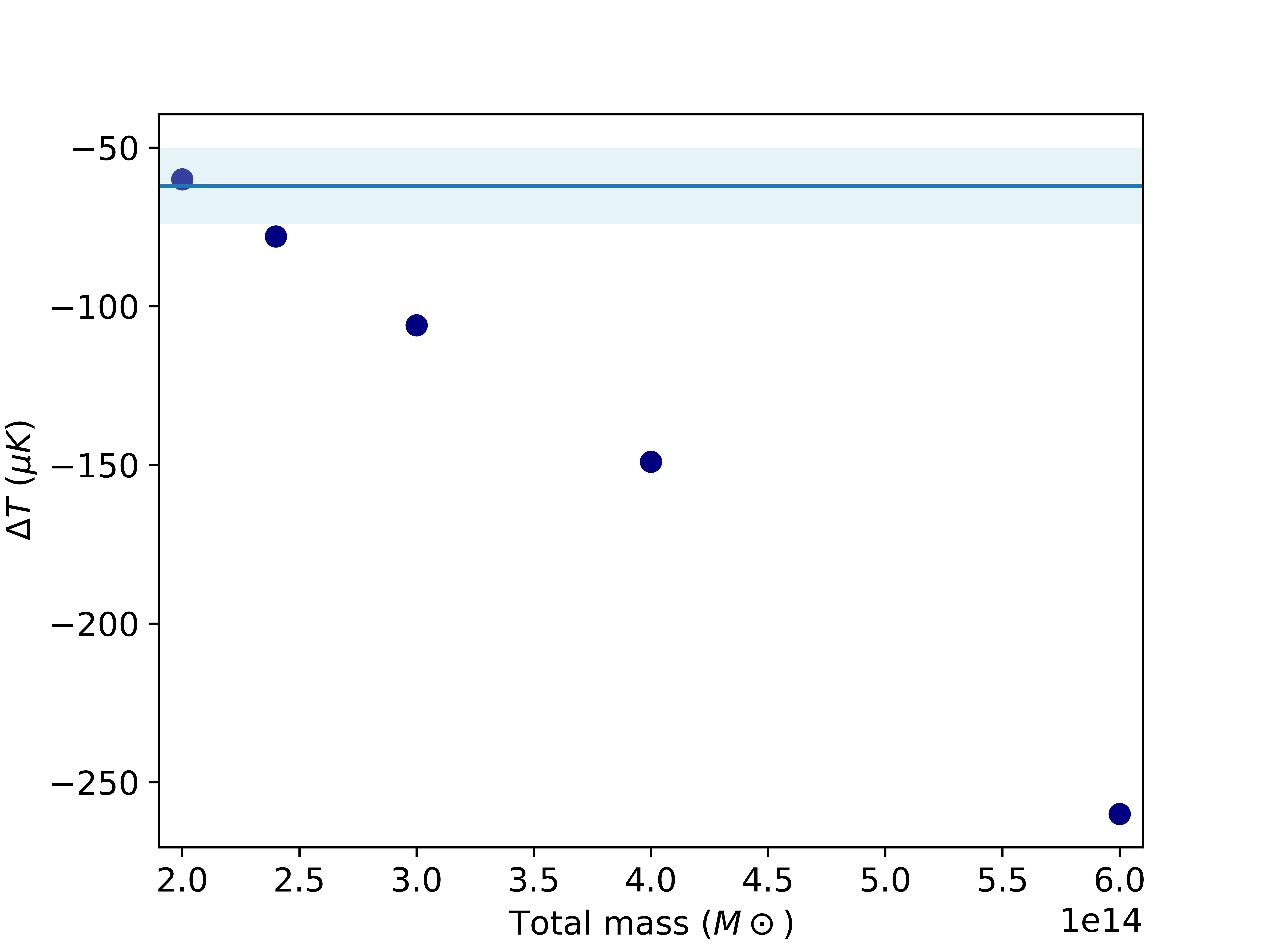}
\caption{ Peak SZ temperature decrement vs cluster total mass. The runs have mass ratio $q=1:1$, ${f}_{\rm gas}=0.05$ and impact parameter $I=0$. The blue line and shaded region represent the observed $\Delta T$ peak and uncertainty respectively. The amplitude of the temperature decrement increases with cluster mass.}
\label{fig:szvstmass}
\end{figure}

Note that for $q=1$, in order to have a single SZ component observed, the clusters would need to be seen in projection directly along the merger axis. However, in that case, no offset would be observed between the SZ and X-ray peaks. Later we consider cases where $q\neq 1$, and where a single SZ component can be observed if the lower-mass component would go undetected. Although the $q=1$ case is ruled out by observations, it is useful to first get a general sense of the impact of total mass $M$ on the evolution of the SZ signal, where a single SZ component would be observed. For each of the runs in Var$_{M}$ and Var$_{c}$ in Table \ref{Table:1} ($q=1$ mergers), the simulation box is projected along the merger axis to give a single SZ peak. For the snapshot with an SZ peak closest to the observations, Fig.\,\ref{fig:szvstmass} shows the peak SZ temperature decrement as a function of $M$. The navy blue line and blue shaded region show the same for the observed system and error range, respectively. As expected, increasing the total mass of the merger system, increases the temperature decrement when the other parameters are fixed. In this case where $q=1$, the total mass of $2\times 10^{14} M_\odot$ produces a peak SZ temperature decrement closest to the observations, whereas higher mass systems produce too large an SZ decrement. This provides us with a starting point for the simulations and depending on the other parameters that we will study in the subsequent sections, the second lowest mass ($2.4 \times 10^{14} M_{\odot}$) may become consistent with the observations. Additionally our preliminary runs confirm that masses above $3\times 10^{14} M_\odot$  for $q \neq 1$ (with masses different enough that we can have a single prominent SZ peak independent of projection axis) cannot produce a peak SZ temperature decrement as low as the observed value.

\begin{figure}
    \centering
\includegraphics[width=0.5\textwidth]{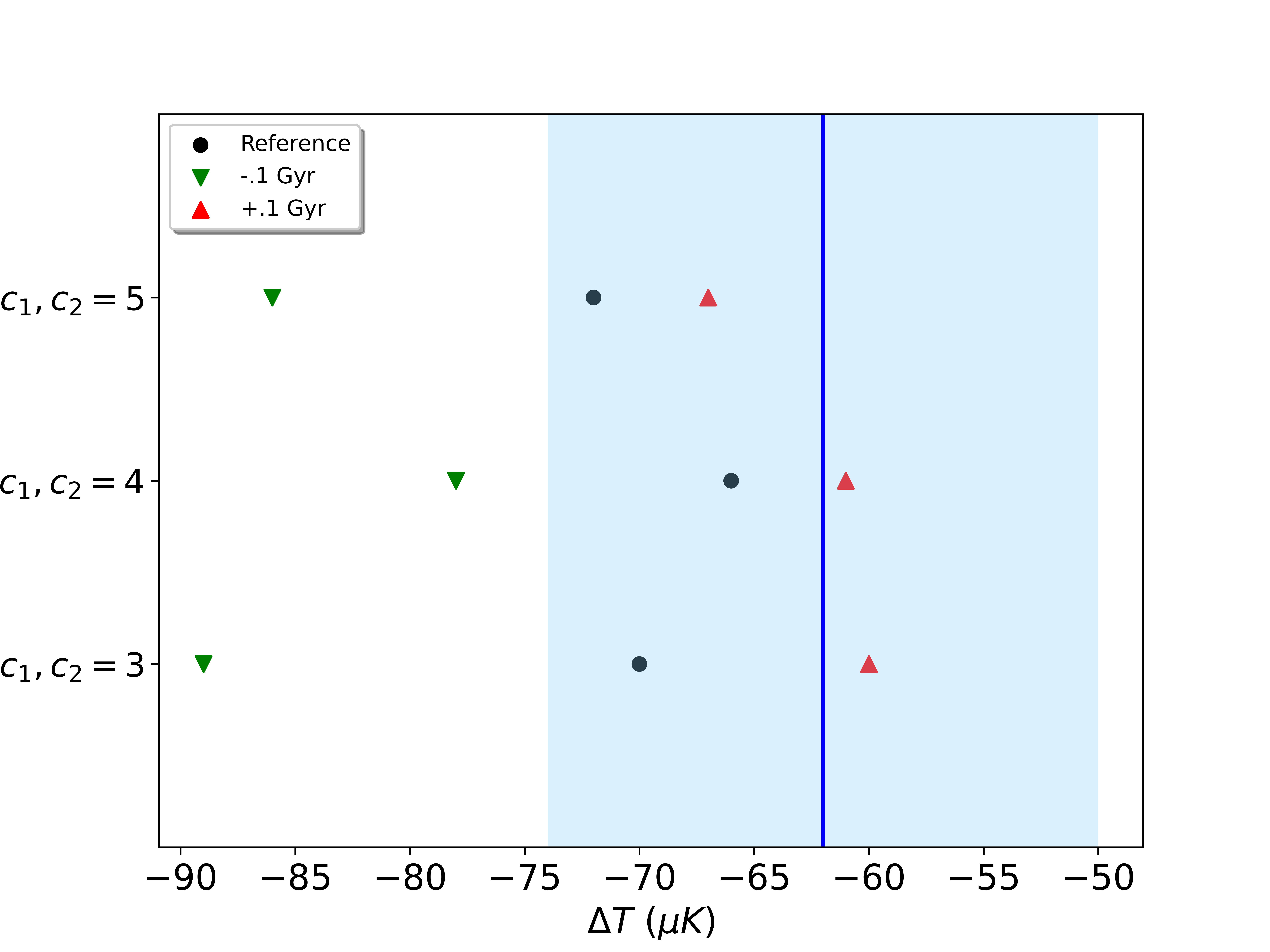}
\caption{
Cluster concentration $c$ vs Peak SZ temperature decrement. All of the runs have mass ratio $q=1:1$, ${f}_{\rm gas}=0.05$ and impact parameter $I=0$. The blue line and shaded region represent the observed $\Delta T$ peak and uncertainty respectively. The solid circle marks the averaged $\Delta T$ around the peak from the earliest map in the simulation runs that falls within the observed range. Concentration is degenerate with temperature decrement, and time from pericentre passage in combination.}
\label{fig:cvsdeltat}
\end{figure}
  
In Fig.\,\ref{fig:cvsdeltat} we vary the initial cluster concentrations in the range $c=3-5$ and show that different concentrations can result in a temperature decrement consistent with the observations. We identify the earliest post-merger simulation snapshot with a $\Delta T$ peak consistent with the observations (solid circles). To assess the impact of snapshot sampling on this measurement, we also identify the snapshots immediately before (green downward arrow) and after (red upward arrow). Concentration is degenerate with SZ temperature decrement and time from pericentre passage in combination. Therefore, we fix $c=5$ in our subsequent analysis, based approximately on the $M-c$ relation \citep{2018ApJ...859...55C}.
 
\begin{figure}
\centering
\includegraphics[width=.5\textwidth]{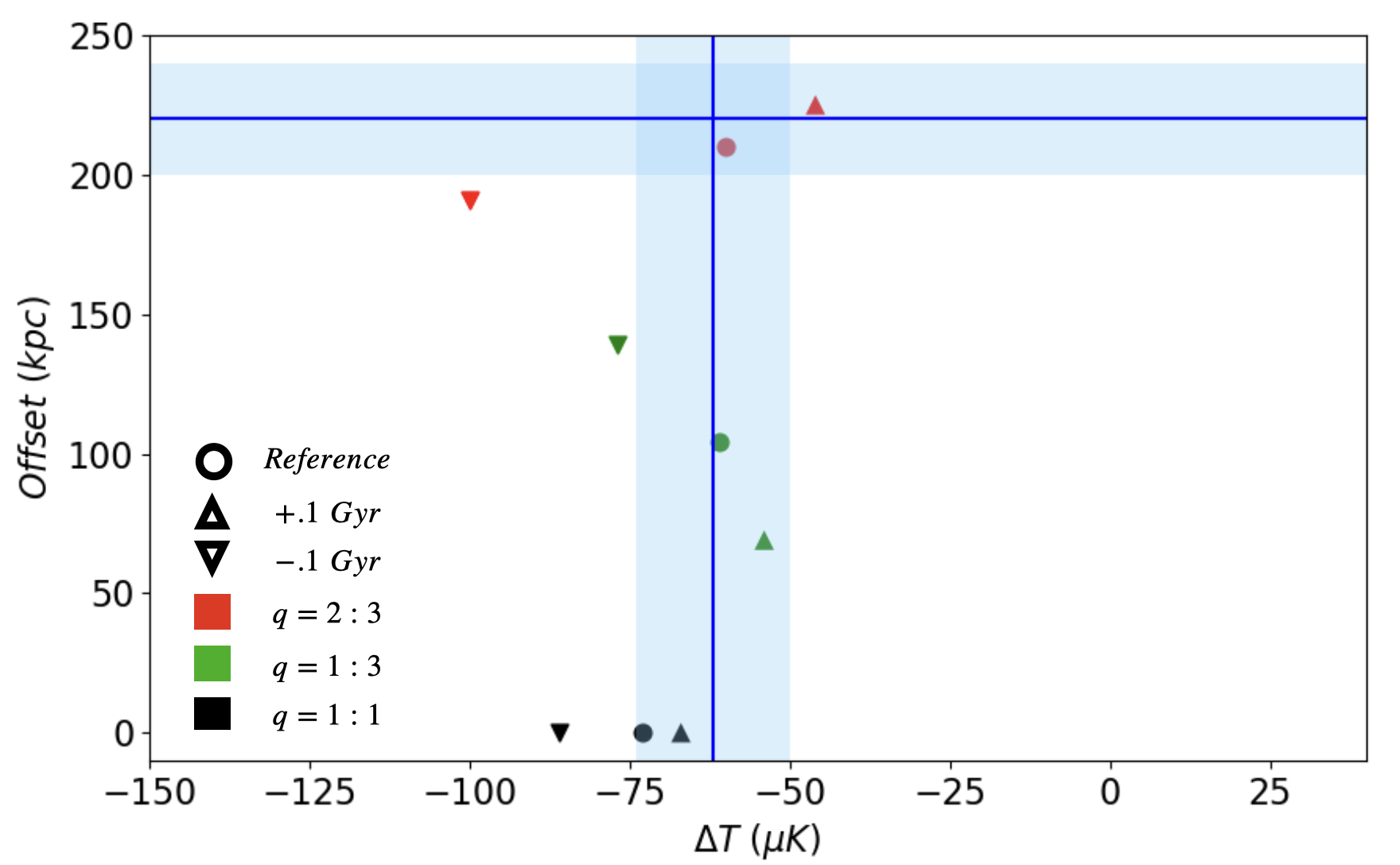}
\caption{Peak SZ temperature decrement vs SZ-X-ray offset as a function of the mass ratio $q$. The vertical (horizontal) navy blue line with the shaded region shows the observed $\Delta T$ (SZ-X-ray offset) and the uncertainty.
The different colours represent SZ peak $\Delta T$ in $\mu K$ and SZ-X-ray offset in kpc for the configurations described in Table \ref{Table:1} Var$_{q}$. The circle represents the first simulation snapshot that falls in the observed range of $\Delta T$ (marked as reference) while the upward (downward) arrows represents the subsequent (previous) snapshot. The offsets are apparent distance between the clusters in the X-Y plane. To match the SZ decrement amplitude and the observed SZ-X-ray offset, a mass ratio of $\sim 2:3$ is needed.}
\label{fig:mr_study}
\end{figure}

Mass ratios, $q$, close to 1:1 are likely to produce the largest offset between SZ and X-ray centres \citep{2014ApJ...796..138Z, 2015ApJ...813..129Z}, in comparison with the smaller mass ratios considered here. In Fig.\,\ref{fig:mr_study}, we show the dependence of the SZ temperature decrement and of the offset on mass
ratio for a total mass of $2\times10^{14}M_\odot$ (Var$_{q}$ in Table \ref{Table:1}). As in Fig.\,\ref{fig:cvsdeltat}, the coloured circles denote the $\Delta T$ and offset measured on the earliest snapshot that is consistent with the observed range, with the downward arrow and upward arrow indicating the previous and subsequent snapshots. For the mass ratios (denoted by different colours) considered, configurations consistent with the observed temperature decrement can be identified, whereas the SZ - X-ray offset is best matched by a mass ratio  $q = 2:3$ and hence we use it in the subsequent studies. Mass ratios $q\sim 1$ are inconsistent with the observations; in that case the requirement of a single prominent SZ peak necessitates a projection along the merger axis which would result in zero offset between the X-ray and SZ peaks in the plane of the sky. 

\begin{figure}
\centering
\includegraphics[width=.5\textwidth]{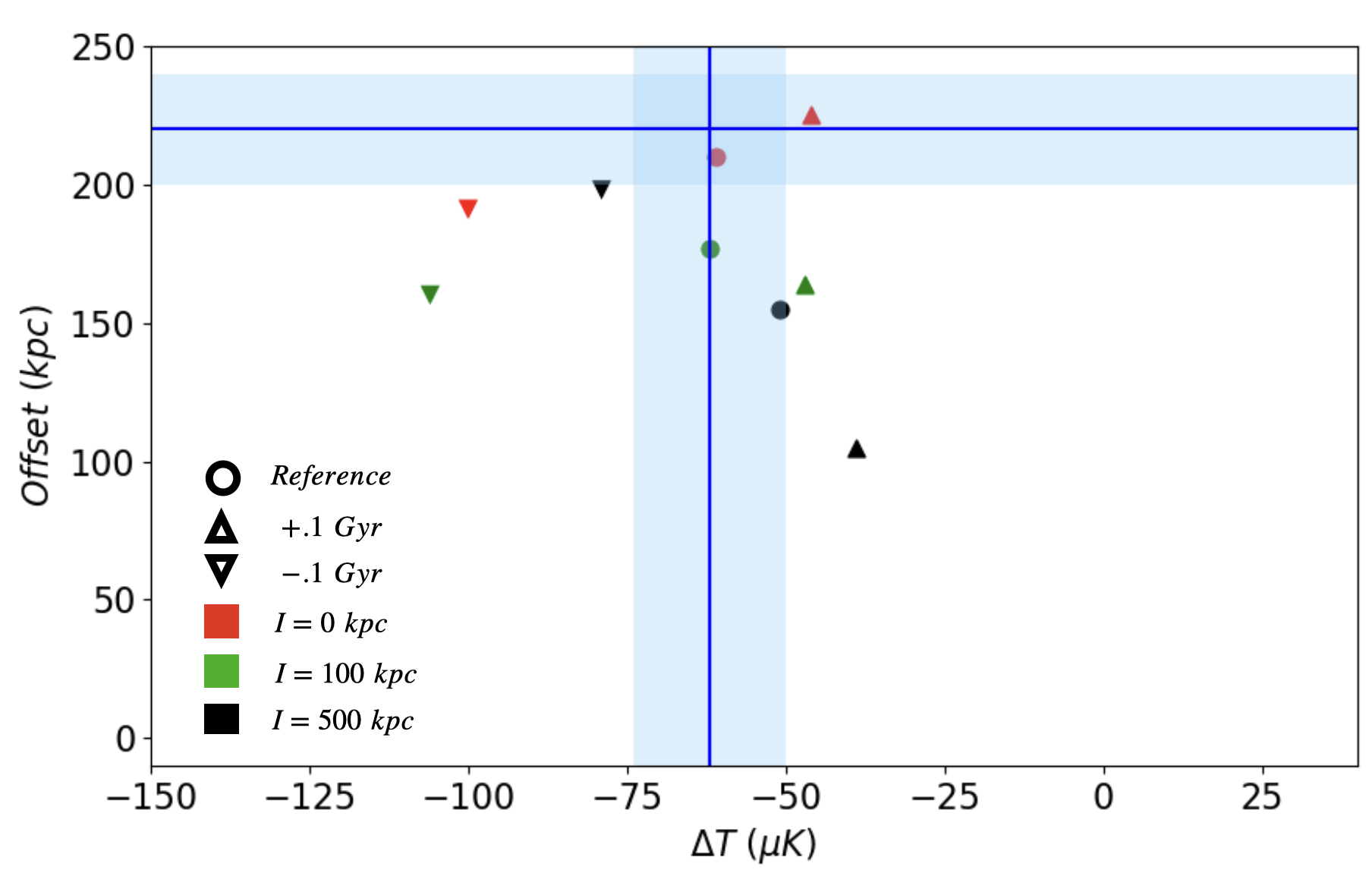}
\caption{ Peak SZ temperature decrement vs SZ-X-ray offset as a function of impact parameter ($I$) for the configurations described in Table \ref{Table:1} Var$_{I}$. The vertical (horizontal) navy blue line and the shaded region show the observed values and the uncertainties respectively.}
\label{fig:ip_study}
\end{figure}

Fig.\,\ref{fig:ip_study}, shows the peak SZ temperature decrement vs observed SZ-X-ray offset as a function of impact parameter. We can identify snapshots consistent with the observed SZ temperature decrement using initial impact parameters in the range $0$ to $500$ kpc studied. Although satisfying the observed offset suggests a smaller impact parameter, we note that snapshots with larger impact parameters can yield offsets roughly consistent with the data.

\begin{figure}
\centering
\includegraphics[width=.5 \textwidth]{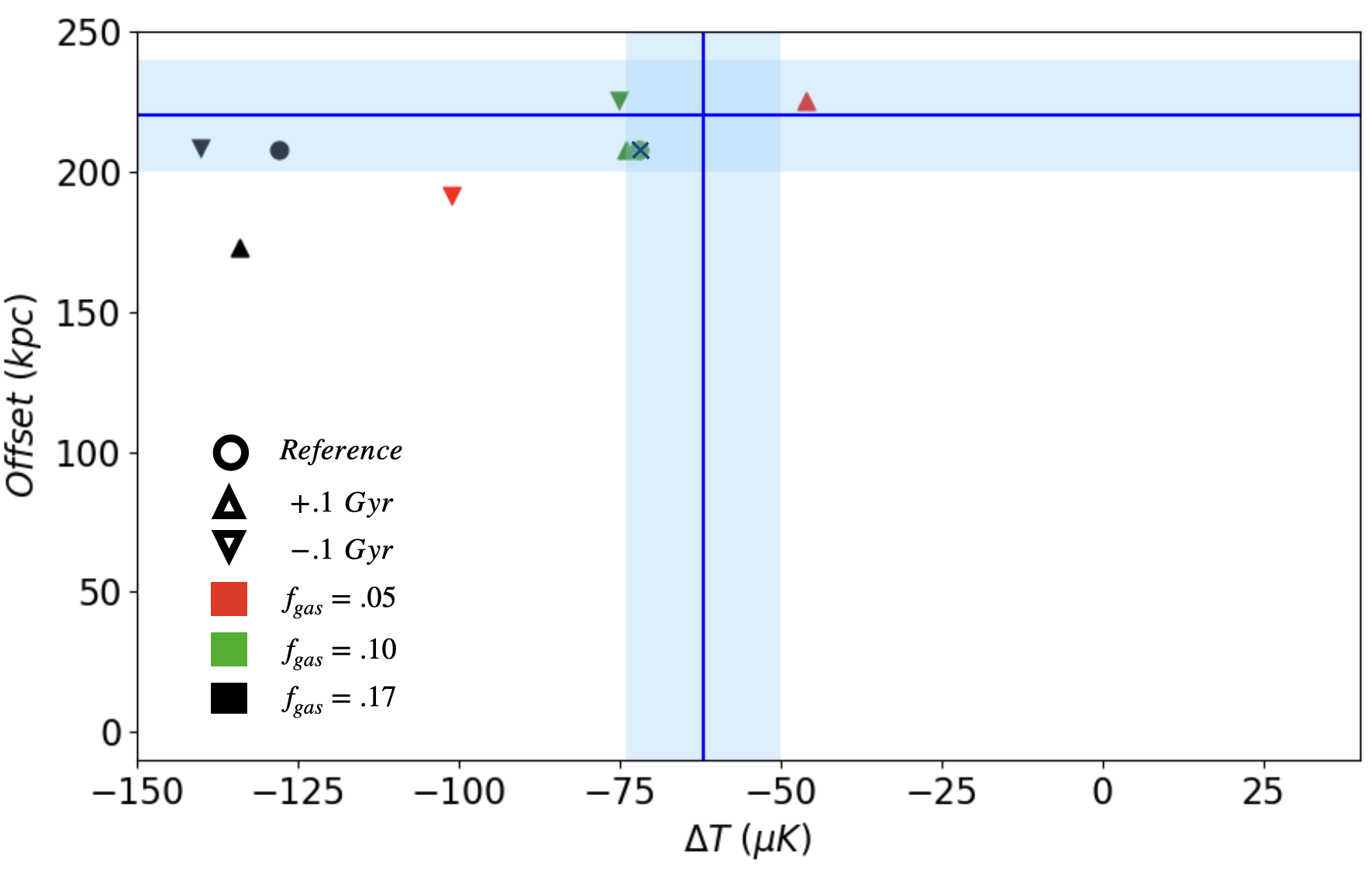}
\caption{Peak SZ temperature decrement vs SZ-X-ray offset for different $f_{gas}$ for the configurations described in Table \ref{Table:1} Var$_{f_{gas}}$. Note that the reference cases for $_{f_{gas}} = 0.05$ and 0.1 are almost coincident and overlap; they are marked with a dark blue X for clarity. The navy blue line and the shaded region show the observed values and the uncertainties. The offset is only mildly sensitive to ${f}_{\rm gas}$. The observed temperature decrement indicates a low ${f}_{\rm gas}$.}
\label{fig:fgas_study}
\end{figure}

\begin{figure*}
    \centering
\includegraphics[width=1.0\textwidth]{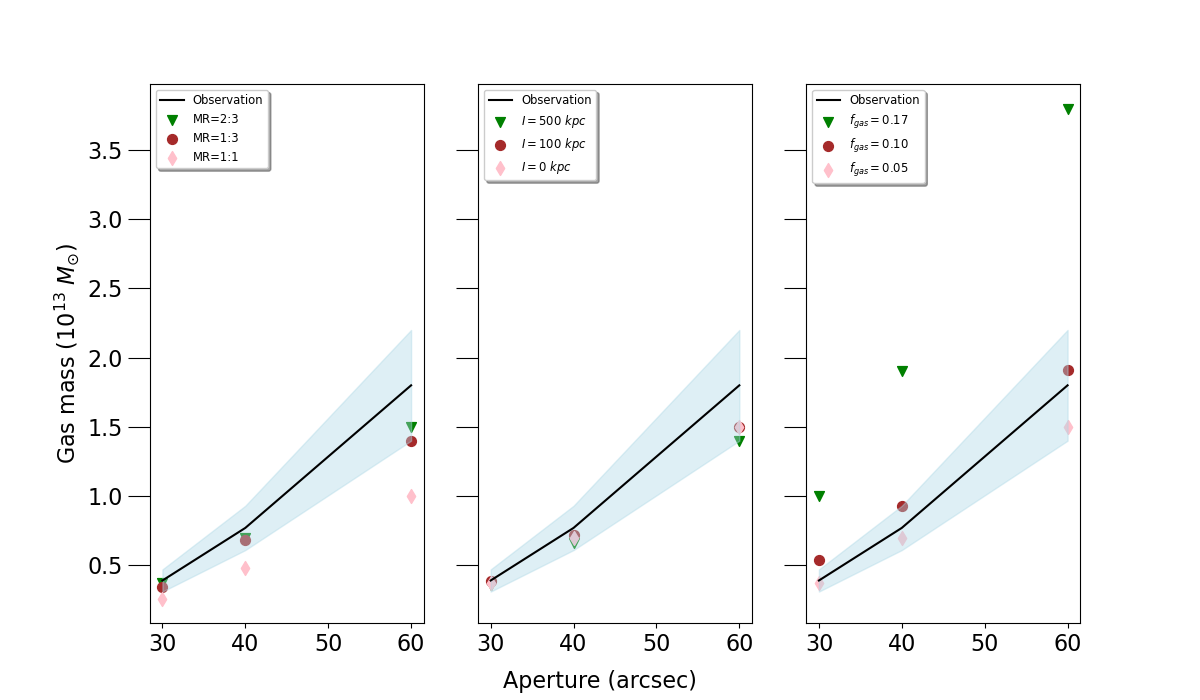}
\caption{Panels show enclosed gas mass vs aperture radius for different mass ratios (left panel), initial impact parameters (central panel), and ${f}_{\rm gas}$ (right panel). The black solid line and the shaded region show the observed gas mass and the uncertainty.}
\label{fig:gmass_all}
\end{figure*}

The SZ and X-ray signal depend on the initial ${f}_{\rm gas}$ of the clusters.  Fig.\, \ref{fig:fgas_study}, shows the SZ-X-ray offset vs peak SZ temperature decrement ( \citet{2023MNRAS.522.4301A}, \citet{2009A&A...507..147A, 2014ApJ...788...51N,  2014A&A...565A.120A}) for different $f_{\rm gas}$.
As expected, increasing ${f}_{\rm gas}$ of the clusters results in increased SZ signal due to increased inverse Compton scattering of the CMB photons. Fig.\,\ref{fig:fgas_study} also strongly suggests that configurations with lower initial ${f}_{\rm gas}$ i.e. $0.05$ - $0.1$ are more consistent with the observations. 

The majority of studies measuring the gas mass fractions of galaxy clusters have been at low redshift 
although recent studies using cosmological simulations find lower ${f_{\rm gas}}$ at higher redshifts, e.g., \citet{2023arXiv230207936A}. Note that some of these studies use different definitions of ${f}_{\rm gas}$, calculated at $r_{500}$ or $r_{2500}$, while we performed the calculation at $r_{200}$. We have measured the gas mass fractions of the most massive clusters extracted at $z\approx 2$ from the Illustris TNG simulations \citep{2017MNRAS.465.3291W,2018MNRAS.473.4077P,2018MNRAS.475..676S}, finding that our conclusion of a lower ${f_{\rm gas}}$ for JKCS041 (in comparison with clusters at $z=0$) is consistent with that sample, albeit that the number of high-redshift massive clusters is limited \citep{2023AAS...24134807S}.

Fig.\,\ref{fig:gmass_all} shows the dependence of (X-ray based) gas mass in different apertures on mass ratio (left panel), impact parameter (central panel) and ${f_{\rm gas}}$ (right panel), as well as the observed values (black line with shading). The gas mass is measured from the first snapshot after core passage to match the observed peak SZ temperature decrement, or the snapshot closest to the observed value (solid blue circles in Figs 5-7). 
At some point after first core passage, all of the impact parameters considered yield gas masses consistent with the observations. Mergers of very comparable mass clusters and gas-rich clusters ($f_{\rm gas}=0.17$) produce gas masses as a function of aperture size that are inconsistent with the observations.

\begin{figure*}
\centering
\includegraphics[width=1.0\textwidth]{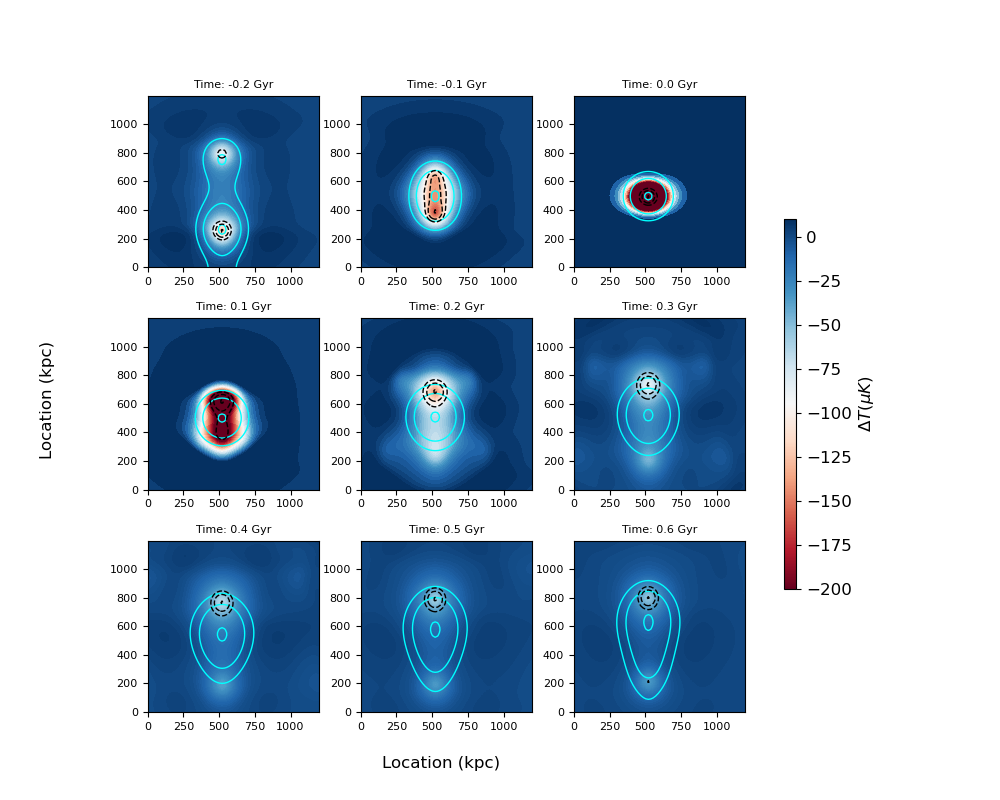}    
    
\caption{Evolution of the SZ signal, shown as $\Delta T$ maps, with projected X-ray emissivity contours, for run $\romannumeral 18$ in Table\,\ref{Table:1}. The SZ map accounts for the transfer function of MUSTANG-2 on GBT, and the X-ray map simulates observations with \textit{Chandra} ACIS-S as in \citet{2009A&A...507..147A} and \citet{2023MNRAS.522.4301A}. The solid cyan (dashed black) contours represent 65 \%, 80 \% and 99 \% levels of projected X-ray emissivity ($\Delta T$). The time relative to first core passage is marked at the top of each panel. The  scenario that is consistent with the observational data is the panel marked 0.3 Gyr. }
\label{fig:szvstime}

\end{figure*}

\begin{figure}
\centering
\includegraphics[width=.48\textwidth]{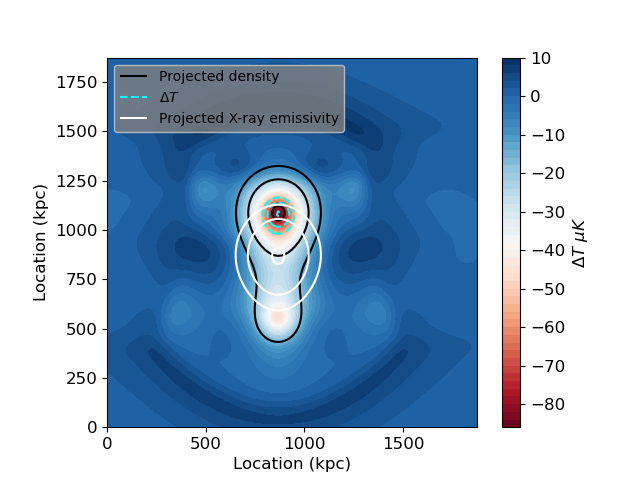}    
    
\caption{Best fit merger model: $\Delta T$ map with projected X-ray emissivity, projected density and $\Delta T$ contour levels for run $\romannumeral 18$ in Table \ref{Table:1}. The map is at time= +0.3 Gyr after first core passage (see Fig.\,\ref{fig:szvstime}) and is the best match to the SZ (averaged $\Delta T$ inside 10'' radius) and X-ray observations and the SZ-X-ray peak offset. The white, black and cyan contours show projected X-ray emissivity, projected density and $\Delta T$ contours at levels 65\%, 80\% and 99\% of the peak value. The averaged $\Delta T$ inside 10'' around the SZ peak is $-61 \mu K$.}
\label{fig:bestfit}

\end{figure}


To summarise our results so far: the amplitude of the SZ temperature decrement is sensitive to the total mass, and to ${f}_{\rm gas}$. The total mass is found to be about $2\times 10^{14} M_{\odot}$, given the remaining parameters. The integrated gas mass inside apertures depends on ${f}_{\rm gas}$, and values below, or of the order of 0.1, are consistent with the observations. The SZ-X-ray offset is sensitive to the mass ratio and to (a lesser extent) the impact parameter. Given also the non-detection of a strong second SZ peak (which rules out roughly equal mass mergers, $q\sim 1$), the mass ratio is found to be about 2:3, given the remaining parameters. None of the observables considered here have a great sensitivity to the initial cluster concentrations.

The 220 kpc offset between the SZ and X-ray peaks is measured in the plane of sky, and the velocity gradient in \citet{2017ApJ...850..203P} and in \citet{2009A&A...507..147A} indicates that the offset also has a component along the line of sight. Thus, 220 kpc is a lower limit on the 3D separation of the SZ and X-ray peaks. In the results above, we focused on configurations where the mergers occur in the plane of the sky, and hence the projected and 3D SZ-X-ray offset are equivalent. 

From our simulations and subsequent analysis, the initial configuration with a total mass of $M=2\times 10^{14}\ M\odot$, mass concentrations of each cluster $c=5$, mass ratio of $q=2:3$, ${f}_{\rm gas}=0.05$ (up to 0.10) and impact parameter of $I=0$ (head on collision) satisfies all of the observational constraints. The evolution of the SZ map of this particular configuration (run $\romannumeral 18$ in Table\,\ref{Table:1}) with time during the merger is shown in Fig.\,\ref{fig:szvstime}. The snapshot at +0.3 Gyr after first core passage is our best fit result to the observations, and is shown in Fig.\,\ref{fig:bestfit}. The compact appearance of the SZ emission associated with the main peak in \citet{2023MNRAS.522.4301A} is reproduced by our simulations.

It is worth noting that this configuration is not exclusive, and that similar configurations with small variations in parameters around this configuration are also consistent with the observations. Generically though, we emphasize that a merger scenario is required to explain all of the observations of JKCS041. Merger scenarios with different initial relative velocities apart from 1100 km/s were also explored (relative velocities of 750 km/s and 1600 km/s), and configurations that match the observations can be identified. All of these velocities were found to be consistent with the observations, with the other parameters fixed as in the best fit configuration. One of the notable differences is that higher initial relative velocities can produce larger SZ-X-ray offsets (1600 km/s produces the largest offset among the velocities we tried). These larger offsets can be reoriented to produce 2D offsets that are consistent with the observations. When exploring other merger configurations (departing from the best fit configuration for the 1100 km/s scenario), in mergers with lower initial velocities (such as 750 km/s), often the maximum 3D SZ-X-ray offset is less than 220 kpc and hence would not be consistent with the observations. For general viewing $\theta$ and $\phi$ angles as indicated on Fig.\,\ref{geometry} (not aligned with the Z-axis where $\theta = 0$ and $\phi = 0$), some merger scenarios can easily be excluded, as with the Z-axis viewing angle. For example, projected SZ-X-ray offset may be too small, or the peak SZ temperature decrement may be too large. However, in general, merger scenarios consistent with the observational constraints can be identified for multiple viewing angles. 
 
\section{Synthetic gravitational lensing analysis}
\label{sec:6}
We now compare the weak gravitational lensing analysis of \citet{kim} with a synthetic lensing analysis of our best-fit merging system. 
Throughout we consider the simulation configuration denoted by "Best Fit" in Table \ref{Table:1}, seen 0.3 Gyr after first core passage. We first describe the creation of synthetic weak lensing catalogues, and then outline how synthetic mass estimates are obtained by analyzing these catalogues.

Weak lensing constraints rely on the observed axis ratios and position angles of distant galaxies being slightly modified by a foreground mass distribution acting as a gravitational lens. In weak lensing analysis, a catalogue of distant galaxies is statistically analyzed, either to map the foreground mass distribution or to obtain a parameterized mass model as considered here. 

\subsection{Generation of synthetic lensing catalogues}
The method we adopt is as follows: for the best fit merger simulation configuration, we obtain the projected density ($\Sigma$) map on a grid, and corresponding lensing convergence ($\kappa$) map, where  $\kappa= \Sigma/\Sigma_{\rm crit}$ and $\Sigma_{\rm crit} = c^{2}D_{s}/4\pi GD_{ds}D_{d}$. The angular diameter distances to the source and lens and the lens-source distance are $D_{s}$, $D_{d}$ and $D_{ds}$ respectively. The redshifts of the lens and sources are taken to be $z=1.8$ and $z_{s}=2.1$ respectively, determining the angular diameter distances in the adopted cosmology. Note that we take $z_{s}=2.1$ for simplicity, consistent with the peak in the redshift distribution in \citet{kim}. We discuss the impact of this assumption in Section \ref{sec:5.2}. A synthetic lensing shear ($\gamma$) map is obtained from the $\kappa$ map using Fourier transform techniques, also yielding a reduced shear map ($g = \gamma/(1-\kappa)$). On the sky, a synthetic distant unlensed galaxy is projected at a location on the $g$ map, sampling a particular value. As outlined below in Eq.\ref{eqsh}, $g$ determines the distorted lensed images of distant galaxy sources. 

In the weak lensing regime, unlensed ($\epsilon^{s}$) and lensed galaxies ($\epsilon$) are described by ellipses, conveniently represented by complex numbers; the modulus of the complex ellipticity is related to the galaxy axis ratio $r=b/a$ via $(1-r)/(1+r)$ and the phase of the complex ellipticity is twice the galaxy position angle.
Similarly, $g$ at any location is represented by a complex number, with a modulus giving the strength, and a phase which is twice the position angle. For each galaxy the lensed and unlensed ellipticities are related via:
\begin{equation}
    \epsilon = \frac{\epsilon^{s} + g}{1 + g^{*}\epsilon^s}\,,
\label{eqsh}    
\end{equation}
where * denotes complex conjugation. Equations relating the intrinsic (unlensed) and lensed shapes of galaxies can be found in \citet{2000A&A...353...41S}, for example. 

In order to generate a synthetic catalogue of lensed galaxies, we adopt a galaxy number density of $\approx$ 87\,arcmin$^{-2}$, randomly distributed and randomly oriented in a square field of view of side 1.5 Mpc for consistency with the data used for the weak lensing analysis of \citet{kim}. The sources are taken to be at a single redshift of $z_{s}=2.1$ (the effective redshift of the \citet{kim} study); the consequences of this assumption are discussed in Section \ref{sec:5.2}. The dispersion of the galaxy ellipticities, quantifying the departure of galaxy shapes from circular, is taken to be $\sigma_{\epsilon^{s}}=0.25$, also in keeping with \citet{kim} and previous studies, e.g., \citep{2000A&A...353...41S}. For all the galaxies in the field, synthetic lensed ellipticities were determined using Equation \ref{eqsh}.
Although synthetic catalogues can also be determined using different sets of random galaxy locations on the sky, here we focus on the impact of galaxy shape noise. 

\subsection{Obtaining weak lensing mass estimates}
\label{sec:5.2}
After creating synthetic lensed galaxy catalogues, we obtained weak lensing mass estimates using the method from \citet{2000A&A...353...41S} and \citet{2001A&A...369....1K}, fitting a single NFW component to the lensed galaxies. 
In keeping with \citet{kim} we use the relation between mass and concentration determined from cosmological simulations, see \citet{2019ApJ...871..168D}, during the fitting. Note that our comparison is with their main mass component. 

\begin{figure}
    \centering
\includegraphics[width=.48\textwidth]{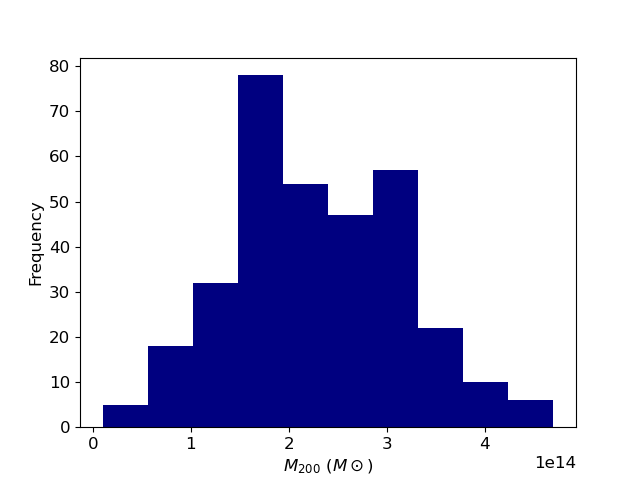}

    \caption{The distribution of recovered $M_{200}$ values for fits to 300 synthetic weak lensing catalogues. The catalogues were created using the best fit merger configuration from run $\romannumeral 18$ in Table \ref{Table:1}. The mean value and standard deviation of the distribution is $2.3 (\pm \, 0.9) \times 10^{14} M_\odot$.}
    \label{fig:lensing}
\end{figure}

Repeating the synthetic catalogue generation and mass recovery 300 times is sufficient to determine mass estimate uncertainties. Fig.\,\ref{fig:lensing} shows the distribution of weak lensing $M_{200}$ fits for synthetic data generated using the best fit configuration. The recovered $M_{200}$ distribution has a mean and standard deviation of $2.3 (\pm \, 0.9) \times 10^{14} M_\odot$, consistent with the observational lensing mass estimate $4.7 (\pm 1.5) \times 10^{14} M_\odot$ (Kim et al, 2023). In our synthetic lensing catalogue generation and analysis we did not account for random distant galaxy positions, or for a source redshift distribution, which would further broaden the distribution of possible $M_{200}$ values. Further, contrary to the case of low redshift clusters, the lensing efficiency for the sources will vary substantially making the distribution of 
redshifts important.

\section{Discussion and Conclusions}
\label{sec:7}
In order to account for the main observable features of JKCS041, we carried out a suite of 
idealized simulations of merging clusters and constructed synthetic SZ and X-ray maps from which synthetic measurements were made. The observed offset between the X-ray and SZ peaks generically requires merger activity, so we restrict our discussion to such systems. We summarize our findings:

\begin{itemize} 
   \item Consistent with results of previous studies, the SZ signal evolves during the merger. After pericentre passage, the peak SZ temperature decrement is lower than that of the component clusters well before pericentre passage. The lower central SZ decrement after pericentre passage is expected since the dark matter separates from the gas leading to decreased gravitational force binding the gas. It is also likely that gas mixing and disturbance of the cool core leads to lower density for the gas distribution after pericentre passage.

   \item The peak SZ temperature decrement and SZ-X-ray offset is only marginally sensitive to concentration parameter $c$. In particular, concentration is degenerate with peak SZ temperature decrement in combination with time from pericentre passage. 
   
   \item Similarly, the initial impact parameter $I$ is, by large, degenerate with peak SZ temperature decrement in combination with time after pericentre passage, while it affects the SZ-X-ray offset observed in the simulations.

   \item The SZ peak is very sensitive to the total mass $M$ and to the initial gas fraction $f_{\rm gas}$ of the system. In particular, higher gas fractions that are typical of present-day clusters are disfavoured. The mass ratio $q$ affects both the SZ peak and the offset. To match the decrement amplitude and the observed SZ-X-ray offset a mass ratio of $\sim 2:3$ is needed.
   
   \item From the suite of cluster merger configurations explored, the observations are consistent with a near head-on cluster merger with $q = 2:3$ seen $\approx 0.3$\,Gyr after first core passage. The cluster components have ${f}_{\rm gas}=0.05$ with summed $M_{200}$ masses $M\approx$$2\times 10^{14}M_\odot$ and (poorly constrained) $c=5$. Values of ${f}_{\rm gas}$ up to $\approx 0.10$ are consistent with the observations of JKCS041.
   
   \item The most important X-ray and SZ observables in constraining the simulation parameters are as follows: the observed SZ peak in JKCS041 requires a low ${f}_{\rm gas}$ in comparison with most observed clusters at $z=0$. The observed offset between the X-ray and SZ peaks requires a merger observed post first pericentre passage. The high redshift of the system ($z=1.8)$, with a Universe of age $\sim 3.66$ Gyr, is indicative of a system seen soon after first pericentre passage rather than at a larger stage of a merger. The single SZ peak and the offset between X-ray and SZ peaks requires a non equal mass merger, where the lower mass object has an undetected SZ peak. 

    \item Synthetic lensing analysis of this merger scenario, fitting a single component NFW halo to the complex system, as done for the real observations by \citet{kim}, yields a mass of 
    $M_{200} = 2.3 (\pm \, 0.9) \times 10^{14}\,M_{\odot}$ consistent with current weak lensing mass estimates from \citet{kim}. Further, the Kim et al. result prohibits significantly lower total mass mergers. 
\end{itemize}

Since the Universe is only 3.6 Gyr old at the redshift of JKCS041, there is a timescale argument that one should consider while modelling the system as a cluster merger. For merger scenarios consistent with JKCS041, sufficient time should have elapsed for cluster creation and subsequent merger events. Idealized merger simulations greatly enhance our ability to interpret the multi-wavelength observable features of a high-redshift system such as JKCS041. Our simulations are idealized, intended to approximately reproduce the observed features, and allowing us to explore a range of cluster merger parameters (e.g., mass, mass ratio, concentration, impact parameter). First, the simulated clusters consist of two spherical cluster-scale dark matter haloes hosting spherical gas distributions, with no substructure on group or galaxy scales. The simulations also do not capture the rich mass accretion history of clusters during the time period in which merger events are simulated. The system is likely in a state of formation where lower mass clusters are in the act of forming a larger cluster, consistent with the growth of clusters seen in cosmological simulations. The high redshift of the system also suggests that the clusters are likely to significantly depart from spherical density profiles. 
 
Our simulations neglect non-thermal contributions to the gas pressure, which arise from turbulence and bulk motions, magnetic fields and cosmic rays.
Studies estimating the non-thermal pressure component in different real clusters at low redshift such as \citet{2018ApJ...861...71S, 2015ApJ...806..207U, 2013MNRAS.428.2241S, 2018A&A...614A...7G} have found that non-thermal pressure could constitute between $5 \%$ to $20 \%$ of the total pressure.\\

In future work, we will complement our idealized simulations by searching for analogues of JKCS041 in very large volume cosmological simulations. This will also allow us to predict the evolution of the system. One measurement that we can make is the evolution of the pressure profile of JKCS041 analogues. The radial pressure profiles of clusters of mass $M_z$ at redshift $z$, $P(r,z,M_z)$ can be related to their pressure profiles if evolved to $z=0$ via
\begin{equation}
P(r,z,M_z) = P(r,z=0,M_{z=0}) E_z^\zeta\,,
\end{equation}
where $r$ is the distance from the cluster centre, $E_z = H(z)/H(z=0)$, and $\zeta\approx$-2 \citep{2021MNRAS.505.5896A}. Our simulations here are idealized, but we will be able to test whether $\zeta\approx$-2 for analogues of this system.
We will also be able to carry out synthetic weak lensing analysis of the more complex systems from cosmological simulations, fitting parameterized mass models with two or more components. As noted in \citet{2023ApJ...945...71L}, the concentrations of galaxy clusters are expected to rise soon after first core passage. Therefore, when assuming a mass-concentration relation during a weak lensing analysis, the mass is expected to be biased high. 

\section*{Acknowledgements}
We acknowledge helpful discussions with Michael Kesden, Regina Jorgenson and Jinhyub Kim. The authors acknowledge HPC resources of Texas Advanced Computing Center (TACC) (http://www.tacc.utexas.edu) and of the GANYMEDE cluster at the University of Texas at Dallas that have contributed to the research results reported within this paper. We acknowledge the support provided by Parvez Jainal, and by the TACC team at The University of Texas at Austin for their help in setting up some of the simulation codes. KS acknowledges support from NSF REU Program NSF-1757321 (PI Regina Jorgenson) and the hospitality of the Maria Mitchell Observatory, and mentoring from Regina Jorgenson.

\section*{Data Availability}
The data arising from this work will be shared on reasonable request to the corresponding authors.

\bibliographystyle{mnras}
\bibliography{draft_v6} 

\appendix

         
    
         
     
         
     



\newpage



\label{app:b}

    \label{fig:Sz_param_vary}


         
    



\bsp	
\label{lastpage}
\end{document}